\documentclass[twocolumn]{revtex4}
\usepackage{amsmath,amsfonts,epsfig}
\usepackage[english]{babel}
\usepackage{graphicx}

\newcommand{\figwidth}{0.45\textwidth}

\begin{document}

\title{Phase diagrams of colloidal spheres with a constant zeta-potential}
\author{Frank Smallenburg$^1$, Niels Boon$^2$, Maarten Kater$^2$, Marjolein Dijkstra$^1$, and Ren\'{e} van Roij$^2$}
\affiliation{$^1$ Soft Condensed Matter, Debye Institute for Nanomaterials Science, Utrecht University, Princetonplein 5, 3584 CC Utrecht, The
Netherlands\\$^2$ Institute for Theoretical Physics, Utrecht University, Leuvenlaan 4, 3584 CE Utrecht, The Netherlands}

\begin{abstract}
We study suspensions of colloidal spheres with a constant zeta-potential within Poisson-Boltzmann theory,
quantifying the discharging of the spheres with increasing colloid density and decreasing salt concentration. We
use the calculated renormalized charge of the colloids to determine their pairwise effective screened-Coulomb
repulsions. Bulk phase diagrams in the colloid concentration-salt concentration representation follow, for various
zeta-potentials, by a mapping onto published fits of phase boundaries of point-Yukawa systems. Although the
resulting phase diagrams do feature face-centered cubic (fcc) and body-centered cubic (bcc) phases, they are
dominated by the (re-entrant) fluid phase due to the colloidal discharging with increasing colloid concentration
and decreasing salt concentration.
\end{abstract}

\maketitle

\section{Introduction}
Charged colloidal particles suspended in a liquid electrolyte are interesting soft-matter systems that have
generated fundamental as well as industrial attention for decades.\cite{hunter} Understanding the stability and
phase behaviour of these systems as a function of colloid concentration and ionic strength is an important theme
in many of these studies. A key role is played by the electrostatic repulsions between the colloidal spheres,
which are not only capable of stabilising suspensions against irreversible aggregation due to attractive Van der
Waals forces,\cite{israel} but are also the driving force for crystallisation,\cite{barrat} provided the surface
charge on the colloids is high enough and the range of the repulsions long enough.\cite{hunter,israel,barrat} The
classic theory that describes the electrostatic repulsions between charged colloidal particles in suspension goes
back to the 1940's, when Derjaguin, Landau, Verwey, and Overbeek (DLVO) found, within linear screening theory,
that suspended spheres repel each other by screened-coulomb (Yukawa) interactions.\cite{dlvo1, dlvo2} The strength of
these repulsions increases with the square of the colloidal charge, and they decay exponentially with
particle-particle separation on the length scale of the Debye screening length of the solvent.\cite{crocker} This
pairwise Yukawa form is a corner stone of colloid science, and can explain a large number of observations.\cite{hunter,israel,barrat} For instance, the experimentally observed crystallisation of charged colloidal
spheres into body-centered cubic (bcc) and face-centered cubic (fcc) phases upon increasing the colloidal packing
fraction at low and high salt concentrations,\cite{sirota,fccbcc1, fccbcc2, fccbcc3} respectively, is in fair agreement with
simulations of Yukawa systems.\cite{rkg,meijer,hamaguchi,hynninen} Interestingly, in these simulation studies, as
well as in many other studies,\cite{alexander,leunissen,dobnikar,zoetekouw} the charge of the colloids is assumed
to be independent of the colloid density and the salt concentration.

The constant-charge assumption was argued to break down, however, in some recent studies where the electrostatic
repulsions were argued to be reduced with increasing colloid concentration. Biesheuvel,\cite{biesheuvel} for
instance, argues that experimental equilibrium sedimentation-diffusion profiles of charged silica spheres in
ethanol at extremely low salt concentrations\cite{rasa} are better fitted by a charge-regulation model than by a
constant-charge model.\cite{rene} More recent evidence for a concentration-dependent colloidal charge stems from
re-entrant melting and re-entrant freezing observations of PMMA spheres in a solvent mixture of {\em cis}-decaline
and cyclohexyl bromide, i.e. the phase sequence upon increasing the colloid concentration is
fluid-crystal-fluid-crystal.\cite{royall} In addition, direct force measurements between a single pair of
colloidal PMMA spheres in hexadecane, a pair that is part of a triplet, and a pair that is part of a multiplet
have very recently revealed a significant reduction of the force with increasing number of neighbouring particles.\cite{merrill} Interestingly, in the three experiments of Refs.\cite{rasa,royall,merrill} the solvent is a
nonpolar medium.

In fact, the experimental findings of Ref.\cite{merrill} could well be interpreted and explained in terms of constant-potential boundary conditions on the colloidal surfaces, rather than the more usual
constant-charge assumption. The present article addresses the consequence of constant-potential boundary conditions for
the packing fraction-salt concentration phase diagram of Yukawa systems by calculating the colloidal charge and
the effective screening length for various zeta-potentials as a function of salt- and colloid concentration.
Perhaps surprisingly, such a study has not yet been performed. In the case of high zeta-potential this requires
nonlinear screening theory, and hence the renormalized rather than the bare colloidal charge determines the
effective screened-Coulomb repulsions between the colloids.\cite{alexander,trizac1, trizac2, trizac3, monovoukas,wette,toyotama,rojas,zoetekouw} For this reason we use the renormalized charge
throughout. We also compare our constant-potential calculations with those of an explicit charge-regulation model,\cite{ninham,chan,behrens,gisler,grunberg,popa2010} and conclude that their results are qualitatively similar,
and even quantitatively if they are considered as a function of the effective screening length.

\section{Model and theory}
We consider $N$ colloidal spheres of radius $a$ in a solvent of volume $V$, temperature $T$, dielectric constant $\epsilon$ and Bjerrum length
$\lambda_B=e^2/\epsilon k_BT$. Here $e$ is the elementary charge and $k_B$ the Boltzmann constant. The colloidal density is denoted by $n=N/V$ and
the packing fraction by $\eta=(4\pi/3)na^3$. The suspension is presumed to be in osmotic contact with a 1:1 electrolyte of Debye length
$\kappa^{-1}$ and total salt concentration $2 \rho_s$. We are interested in suspensions of charged colloids of which the surface (zeta) potential
$\psi_0$ rather than the charge $Ze$ is fixed. We will show that this constant-potential condition mimics charge-regulation on the colloidal
surfaces fairly accurately. The first goal of this article is to calculate $Z$ as a function of $\eta$ for fixed dimensionless combinations
$\kappa a$, $a/\lambda_B$, and $\phi_0\equiv e\psi_0/k_BT$. This result will then be used to quantify the effective Yukawa interactions between
pairs of colloids, and hence the phase boundaries between fluid, face-centered cubic (fcc) and body-centered cubic (bcc) crystalline phases.

In the actual suspension of constant-potential colloidal spheres, the charge distribution of each of the $N$ colloids will be distributed
heterogeneously over its surface due to the proximity of other colloids in some directions. This leads to a tremendously complex many-body problem
that we simplify here by {\em assuming} a spherically symmetric environment for each colloid, which is nevertheless expected to describe the
average electrostatic properties realistically. Below we will calculate the electrostatic potential $\psi(r)$ at a radial distance $r$ from a
charged colloidal sphere at a given zeta-potential $\psi_0$, i.e. at a given value $\psi(a)=\psi_0$. The colloidal charge  $Ze$ then follows from
Gauss' law
\begin{equation}
\psi'(a)=-\frac{Ze}{\epsilon a^2}, \label{gauss}
\end{equation}
where a prime denotes a radial derivative.

We first consider a single colloid in the center of a Wigner-Seitz cell of radius $R$, such that the cell volume equals the volume per particle,
$(4\pi/3)R^3=V/N$, which implies $R=a\eta^{-1/3}$. The radial coordinate of the cell is called $r$. We write the ionic density profiles for
$r\in(a,R)$ as Boltzmann distributions $\rho_{\pm}(r)=\rho_s\exp(\mp\phi(r))$, with $\phi(r)=e\psi(r)/k_BT$ the dimensionless electrostatic
potential. Together with the Poisson equation $\nabla^2\phi=-4\pi\lambda_B(\rho_+(r)-\rho_-(r))$, this gives rise to the radially symmetric
PB-equation and boundary conditions (BC's)
\begin{eqnarray}
\phi''(r)+\frac{2}{r}\phi'(r)&=&\kappa^2\sinh\phi(r), \hspace{4mm}r\in(a,R);\label{pb1}\\
\phi(a)&=&\phi_0; \label{bc1}\\
\phi'(R)&=&0,\label{bc2}
\end{eqnarray}
where a prime denotes a derivative with respect to $r$. Note that BC (\ref{bc2}) implies charge neutrality of the cell. Once the solution
$\phi(r)$ is found for given $\eta$, $\kappa a$, and $\phi_0$, e.g. numerically on a radial grid, the colloidal charge $Z$ follows from
Eq.(\ref{gauss}), which we rewrite in dimensionless form as
\begin{equation}
\frac{Z \lambda_B}{a}=-a\phi'(a). \label{Zws}
\end{equation}
From the numerical solutions that we will present below it turns out that $Z$ decreases monotonically from a finite asymptotic low-$\eta$
(large-$R$) value $Z_0$ to essentially 0 at $\eta\simeq 1$ (or $R\simeq a$).

Within linear-screening theory at low packing fraction, where $\sinh\phi\simeq\phi$, the potential distribution can be solved for analytically,
yielding $\phi(r)=\phi_0 a\exp[-\kappa (r-a)]/r$, such that the colloidal charge takes the asymptotic low-$\eta$ and low-$\phi_0$ value
\begin{equation}
\frac{Z_0\lambda_B}{a}=(1+\kappa a)\phi_0. \label{Z0}
\end{equation}
In the appendix we show that the discharging effect with increasing $\eta$, as found from the nonlinear screening theory discussed above, can also
be approximated within linear screening theory, yielding
\begin{equation}
Z(\eta,\kappa a)=\frac{Z_0}{1+\eta/\eta^*}, \hspace{1cm} \eta^*=\frac{(\kappa a)^2}{3(1+\kappa a)}, \label{Zeta}
\end{equation}
where $\eta^*$ is a crossover packing fraction at which the colloidal charge has decayed to half its dilute-limit value $Z_0$ given in
Eq.(\ref{Z0}). For typical numbers of experimental interest, e.g. $a/\lambda_B=100$ and $\kappa a=0.25$, we then find $Z_0=125\phi_0$ and
$\eta^*=0.017$. With $\phi_0\simeq 1-2$, which corresponds to a surface potential of 25-50mV, we should expect a few hundred charges in the dilute
limit and a significant charge reduction for $\eta\gtrsim 10^{-2}$.

The constant-potential boundary condition that we employ here is supposed to mimic charge-regulation on the colloidal surface through an
association-dissociation equilibrium of chargeable groups on the surface. Here we consider, as a typical example, the reaction SA$\rightleftharpoons$
S$^+$+A$^-$, where a neutral surface group SA dissociates into a positively charged surface group S$^+$ and a released anion A$^-$. The chemistry
of such a reaction can be characterised by a reaction constant $K$ such that [S$^+$][A$^-$]/[SA]=$K$, where the square brackets indicate
concentrations in the vicinity of the surface where the reaction takes place. If we now realise that $Z\propto[$S$^+$], we find for the usual case
where [S$^+]\ll$[SA] that $Z\propto1/$[A$^-$]. For the case that the released anion is of the same species as the anion in the reservoir, such
that [A$^-]=\rho_s\exp[\phi(a)]$, we thus have
\begin{eqnarray} \label{ZY} Z=z\exp(-\phi(a)),
\end{eqnarray}
where the prefactor $z$, which is a measure for the surface chargeability,\cite{niels} accounts for the
chemistry, the surface-site areal density, and the total area of the surface between the colloidal particle and
the electrolyte solution. Note that Eq.(\ref{ZY}) relates the (yet unknown) colloidal charge $Z$ to the (yet
unknown) zeta-potential $\phi(a)$, for a given $z$. A closed set of equations for charge-regulated colloids is
obtained by combining the PB equation (\ref{pb1}) with BC (\ref{bc2}) at the boundary of a Wigner-Seitz cell of
radius $R$, with BC (\ref{bc1}) replaced by
\begin{eqnarray}
a\phi'(a)=-\frac{\lambda_B z}{a}\exp(-\phi(a)), \end{eqnarray} for some given chargeability $z$. The resulting
solution $\phi(r)$ gives the zeta-potential $\phi(a)$ as well as the colloidal charge $Z$ using Eq.(\ref{ZY}).
When comparing the constant-potential model with the ionic association-dissociation model, we will tune the
chargeability $z$ such that the low-$\eta$ results for $Z$ coincide for both models.

It is well known that nonlinear screening effects, in particular counterion condensation in the vicinity of a
highly charged colloidal surface, reduce the effective colloidal charge that dictates the screened-Coulomb
interactions between the colloids.\cite{alexander,trizac1, trizac2, trizac3, zoetekouw,palberg} The so-called renormalized colloidal
charge, $Z^*$,\cite{lowen} can be calculated from the electrostatic potential $\phi(r)$ as obtained from the nonlinear PB
equation by matching the numerically obtained solution at the edge of the cell to the analytically known solution
of a suitably linearized problem. By extrapolating the solution of the linearized problem to the colloidal surface
at $r=a$, one obtains the effective charge by evaluating the derivative at $r=a$ using Eq.(\ref{Zws}). Following
Trizac {\it et al.},\cite{trizac03} the renormalized charge $Z^*$ can be written as
\begin{eqnarray}
 \frac{Z^*\lambda_B}{a} &=& -\frac{\tanh \phi_D}{\bar{\kappa} a}\left( (\bar{\kappa}^2 a R-1) \sinh[\bar{\kappa}(R-a)]\right. +\nonumber\\
& & \left.\bar{\kappa}(R-a) \cosh[\bar{\kappa}(R-a)] \right),
\end{eqnarray}
where the `Donnan' potential is defined as $\phi_D\equiv\phi(R)$, i.e. the numerically found potential at the
boundary of the cell, and where the effective inverse screening length is
\begin{eqnarray} \label{kappabar}
\bar{\kappa} = \kappa \sqrt{\cosh\phi_D}.
\end{eqnarray}
Note that $Z^*$ and $\bar{\kappa}$ can be calculated for the constant-potential as well as the association-dissociation model in a Wigner-Seitz
cell.

\section{Effective charge and screening length}
\begin{center}
\begin{figure}
\includegraphics[width=\figwidth]{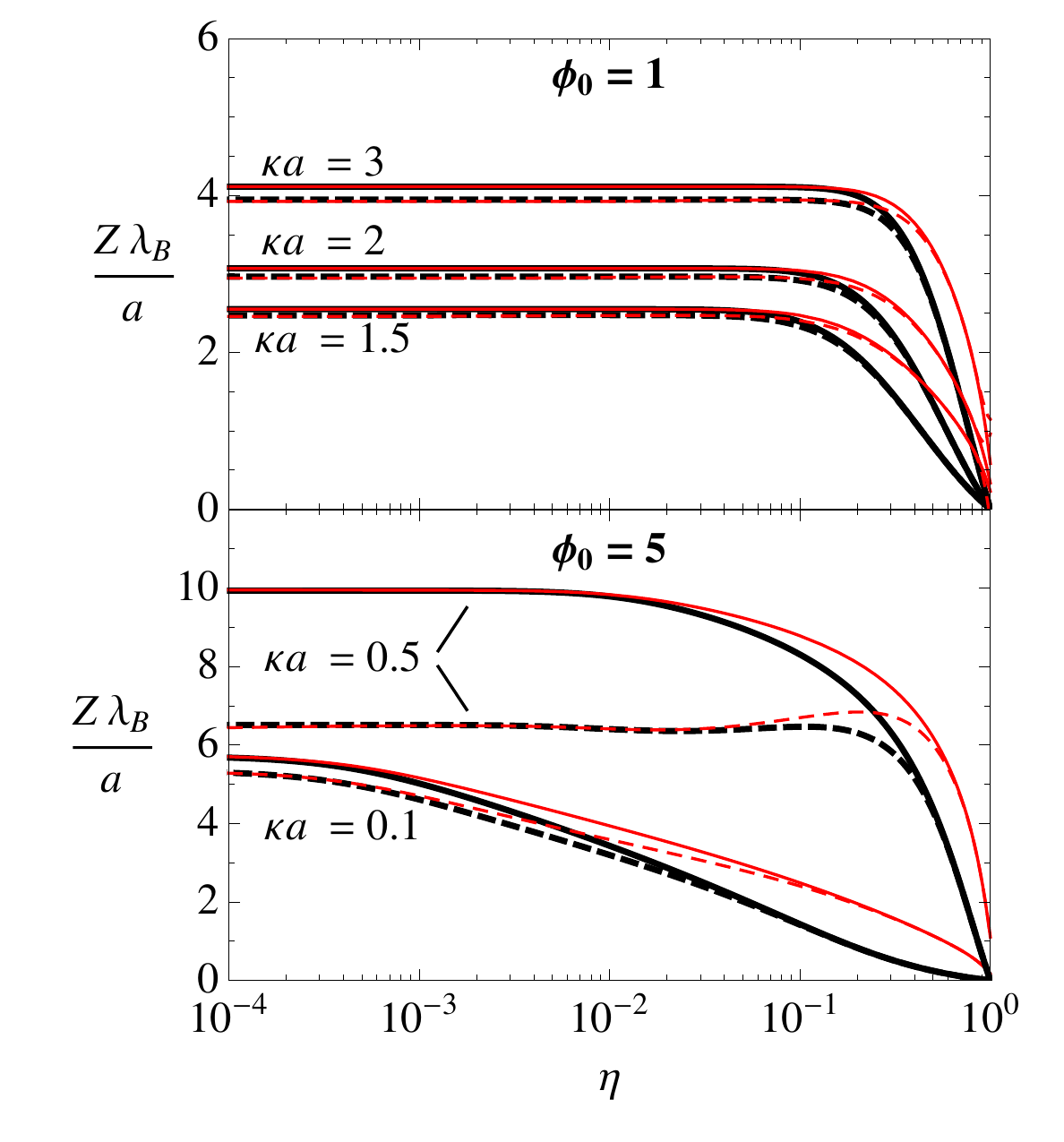}
\caption{The bare colloidal charge $Z$ (continuous black curves) and the renormalized charge $Z^*$ (dashed black curves), both in units of
$a/\lambda_B$ (see text), as a function of the colloidal packing fraction $\eta$ for several screening parameters $\kappa a$, for constant surface
potentials (a) $\phi_0=1$ and (b) $\phi_0=5$. The red curves denote $Z$ and $Z^*$ as obtained from the association-dissociation model, with the
chargeability $z$ chosen such that the surface potential in the dilute limit $\eta\rightarrow 0$ equals $\phi_0$. } \label{chargeplots}
\end{figure}
\end{center}

For both the constant surface potential (CSP) and the association-dissociation (AD) model discussed above we calculated the bare colloidal charge
$Z$, the effective (renormalized) charge $Z^*$, and the effective inverse screening length $\bar{\kappa}$ in the geometry of Wigner-Seitz cells.
In Fig.\ref{chargeplots} we show $Z\lambda_B/a$ (full curves) and $Z^*\lambda_B/a$ (dashed curves) as a function of packing fraction $\eta$, for
two screening constants for both the CSP model (black curves) and the AD model (red curves), in (a) for fixed zeta-potential $\phi_0=1$ and in (b)
for $\phi_0=5$. In all cases the chargeability parameter $z$ of the AD model is chosen such as to agree with the CSP model in the low-density
limit $\eta\rightarrow 0$. The semi-quantitative agreement between the red and black curves for equal $\kappa a$ is indicative of the reasonable
description of charge-regulation by constant-potential BC's, certainly at low $\eta$. At higher $\eta$ the charges predicted by the AD model
exceed those of the CSP model somewhat, which should not come as a surprise since the former interpolates between the constant-charge and the
constant-potential model. The close agreement between $Z$ and $Z^*$ for all $\kappa a$ at $\phi_0=1$ in Fig.\ref{chargeplots}(a) is also to be
expected, since $\phi_0=1$ is not far into the nonlinear regime. By contrast, deep in the nonlinear regime of $\phi_0=5$, as shown in
Fig.\ref{chargeplots}(b), there is a significant charge renormalisation effect such that $Z^*<Z$ by a factor of about 1.2 and 1.5 for $\kappa
a=0.1$ and $\kappa a=0.5$, respectively. The merging of the red and the black curves at high-$\eta$ in Fig.\ref{chargeplots}(b) is due to the
reduction of the charge into the linear-screening regime such that $Z=Z^*$. The increase of $Z^*$ with $\kappa$, as observed in both
Fig.\ref{chargeplots}(a) and (b), is in line with well-known charge-renormalisation results,\cite{alexander,trizac1, trizac2, trizac3,trizac03,zoetekouw} and with Eq. \ref{Z0}.

%%%%%%%% FIG   %%%%%%%%%%%%%%%%%%%%%%%%%%%%%%%%%%%%%%%%%%%%%%%%%%%%%%%%%
%%%%%%%% FIG   %%%%%%%%%%%%%%%%%%%%%%%%%%%%%%%%%%%%%%%%%%%%%%%%%%%%%%%%%

%%%%%%%% FIG   %%%%%%%%%%%%%%%%%%%%%%%%%%%%%%%%%%%%%%%%%%%%%%%%%%%%%%%%%
\begin{center}
\begin{figure}
\includegraphics[width=\figwidth]{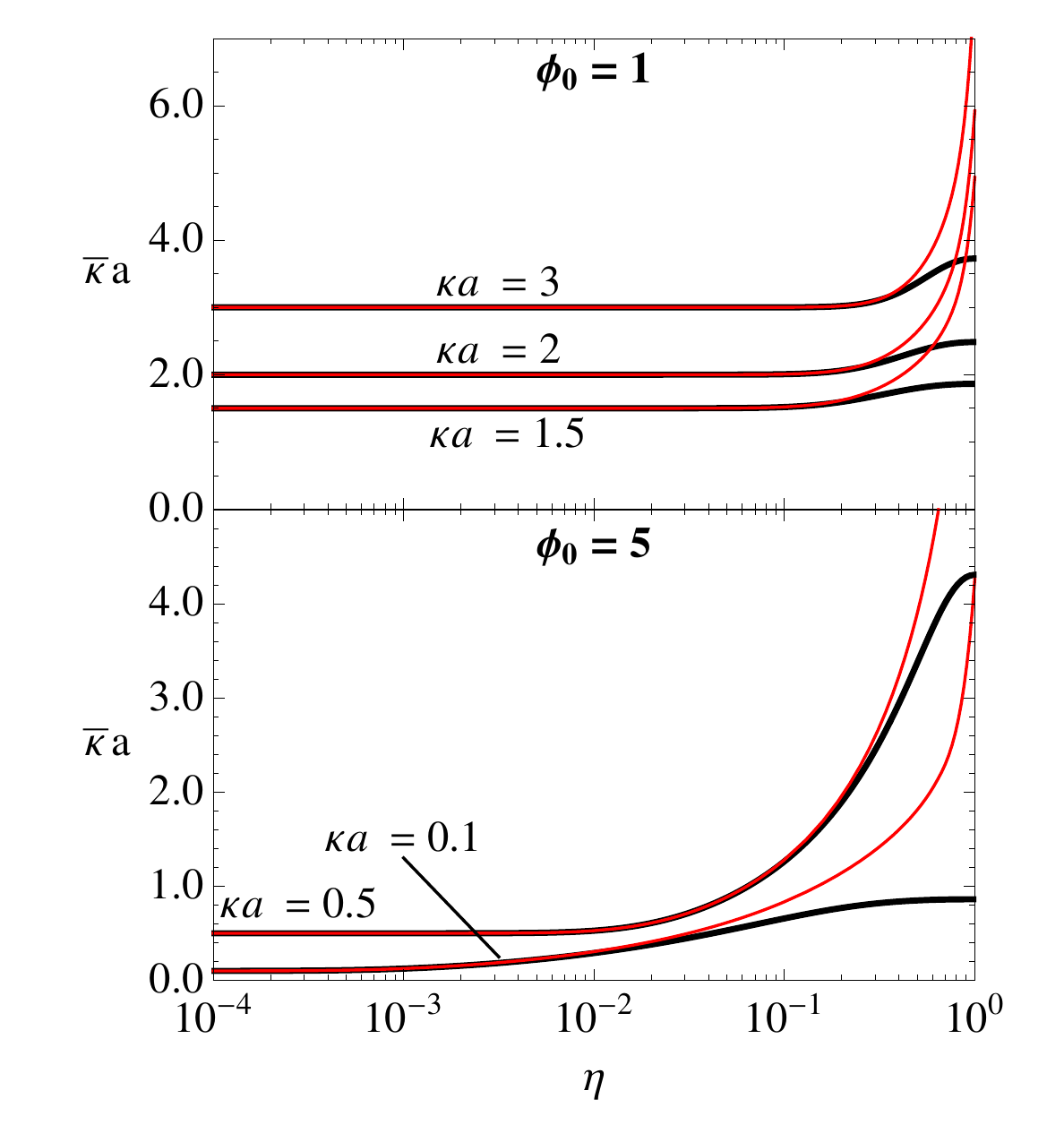}
\caption{The effective inverse screening length $\bar{\kappa}$ as a function of the packing fraction $\eta$ for several reservoir screening
parameters $\kappa a$, for constant surface potentials (a) $\phi_0=1$ and (b) $\phi_0=5$ as represented by the black curves. The red curves denote
$\bar{\kappa}$ as obtained from the association-dissociation model, with the chargeability $z$ chosen such that the surface potential in the
dilute limit $\eta\rightarrow 0$ equals $\phi_0$. Note that $\bar{\kappa}=\kappa$ in all cases for $\eta\rightarrow 0$.} \label{kappaplots}
\end{figure}
\end{center}
%%%%%%%% FIG   %%%%%%%%%%%%%%%%%%%%%%%%%%%%%%%%%%%%%%%%%%%%%%%%%%%%%%%%%

In Fig.\ref{kappaplots}(a) and (b) we plot, for the same zeta-potentials as in Fig.1(a) and (b), the effective screening parameter $\bar{\kappa}$
as a function of $\eta$ for several reservoir screening constants $\kappa$. At low enough $\eta$, where $\kappa R\gg 1$, the two screening
constants are indistinguishable from each other in all cases. The reason is that the cell is then large enough for the potential to decay to
essentially zero at $r=R$, such that the asymptotic decay of $\phi(r)$ is governed completely by the screening constant $\kappa$ of the background
(reservoir) salt concentration. At larger $\eta$, and hence smaller cells, $\phi(R)$ is no longer vanishingly small and the ion concentrations
$\rho_{\pm}(R)$ at $r=R$ deviate considerably from the ionic reservoir concentration $\rho_s$. This larger ionic concentration at the cell
boundary, which represents an enhanced ion concentration in between the colloidal particles in the true many-body system, leads to a larger
effective screening constant $\bar{\kappa}$ with increasing $\eta$ at a fixed $\kappa$, as is shown in Fig.\ref{kappaplots}(a) and (b). Given that
larger charges are obtained  in the AD model than in the CSP model at high $\eta$, the number of counterions in the cell, and hence
$\bar{\kappa}$, is also larger in the AD model.

\section{Effective interactions and phase diagrams}
Once the effective colloidal charge $Z^*$ and the effective screening length $\bar{\kappa}^{-1}$ have been
determined from the numerical solution of the PB equation in a Wigner-Seitz cell, either for constant-potential or
association-dissociation boundary conditions, the effective interactions $u(r)$ between a pair of colloidal
particles separated by a distance $r$ follows, assuming DLVO theory, as
\begin{eqnarray} \label{eq:dlvo}
\frac{u(r)}{k_BT} &=& \left\{
\begin{array}{ll}\displaystyle \infty, &r<2a;\\
\displaystyle \lambda_B\left(\frac{Z^*\exp(\bar{\kappa}a)}{1+\bar{\kappa} a}\right)^2\frac{\exp(-\bar{\kappa} r)}{r}, & r>2a,
\end{array}\right.
\end{eqnarray}
where we include a short-range hard-core repulsion for overlapping colloids and ignore Van der Waals forces (which is justified for index-matched
particles). One could use Eq.(\ref{eq:dlvo}) for the pair interaction to simulate (or otherwise calculate) properties of the suspension in a given
state-point, e.g. whether the system is in a fluid or crystalline state. We restrict our attention here to the limiting case in which the
colloidal particles are sufficiently highly charged and/or sufficiently weakly screened, that the pair-potential at contact satisfies $u(2a)\gg
k_BT$, thereby effectively preventing direct particle-particle contact. In this limit the suspension can be effectively regarded as a point-Yukawa
system that can be completely characterised by only {\em two} dimensionless parameters $U$ and $\lambda$ for the strength and the range of the
interactions, respectively. They are defined as
\begin{eqnarray}
 U      &=& \left(\frac{Z^*\exp(\bar{\kappa} a)}{1+\bar{\kappa} a}\right)^2 \frac{\lambda_B}{a}  \left(\frac{3\eta}{4 \pi}\right)^{1/3}\label{U}\\
 \lambda   &=& \bar{\kappa} a \left(\frac{3\eta}{4 \pi}\right)^{-1/3}\label{lambda},
\end{eqnarray}
such that the point-Yukawa interaction potential of interest, in units of $k_BT$, reads  $U \exp(-\lambda x)/x$ with $x=r(N/V)^{1/3}$ the particle
separation in units of the typical particle spacing. Note that {\em three} dimensionless parameters would have been needed if hard-core contact
was not a low Boltzmann-weight configuration, e.g. then the contact-potential $\beta u(2a)$ (i.e. the inverse temperature), the dimensionless
screening parameter $\kappa a$, and the packing fraction $\eta$ would be a natural choice. The mapping of the phase diagram of the point-Yukawa
system onto hard-core Yukawa systems has been tested and verified explicitly by computer simulation.\cite{hynninen}

%%%%%%%% FIG   %%%%%%%%%%%%%%%%%%%%%%%%%%%%%%%%%%%%%%%%%%%%%%%%%%%%%%%%%
\begin{center}
\begin{figure*}
\includegraphics[width=\figwidth]{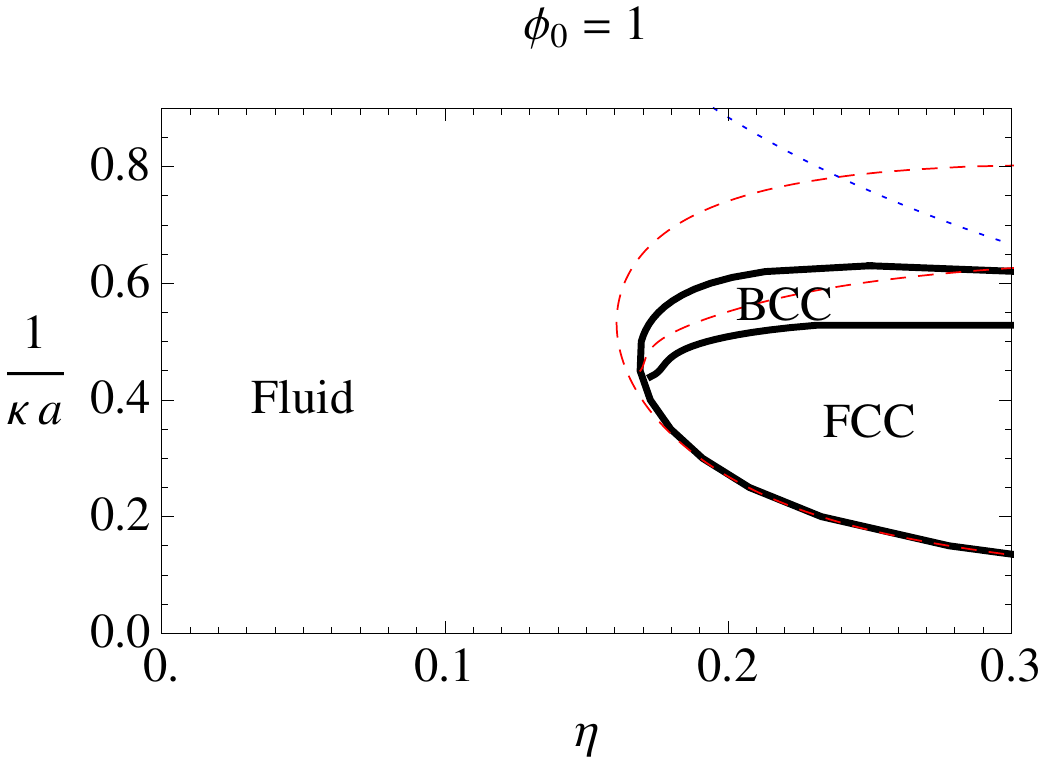}
\includegraphics[width=\figwidth]{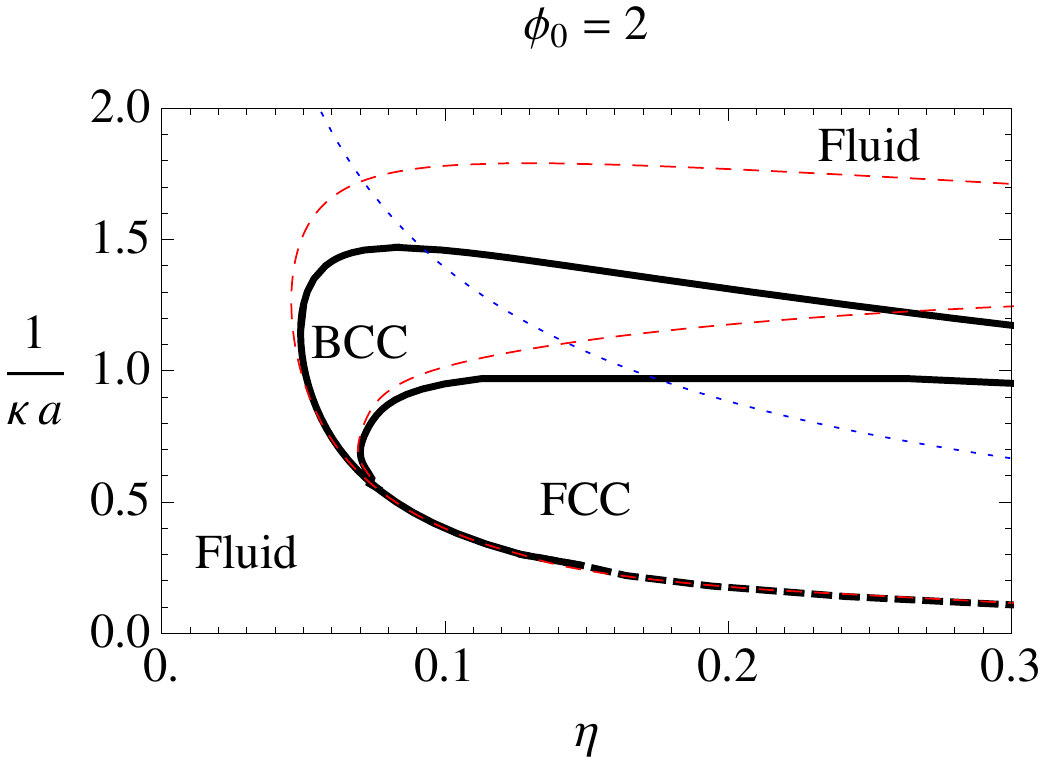}
\includegraphics[width=\figwidth]{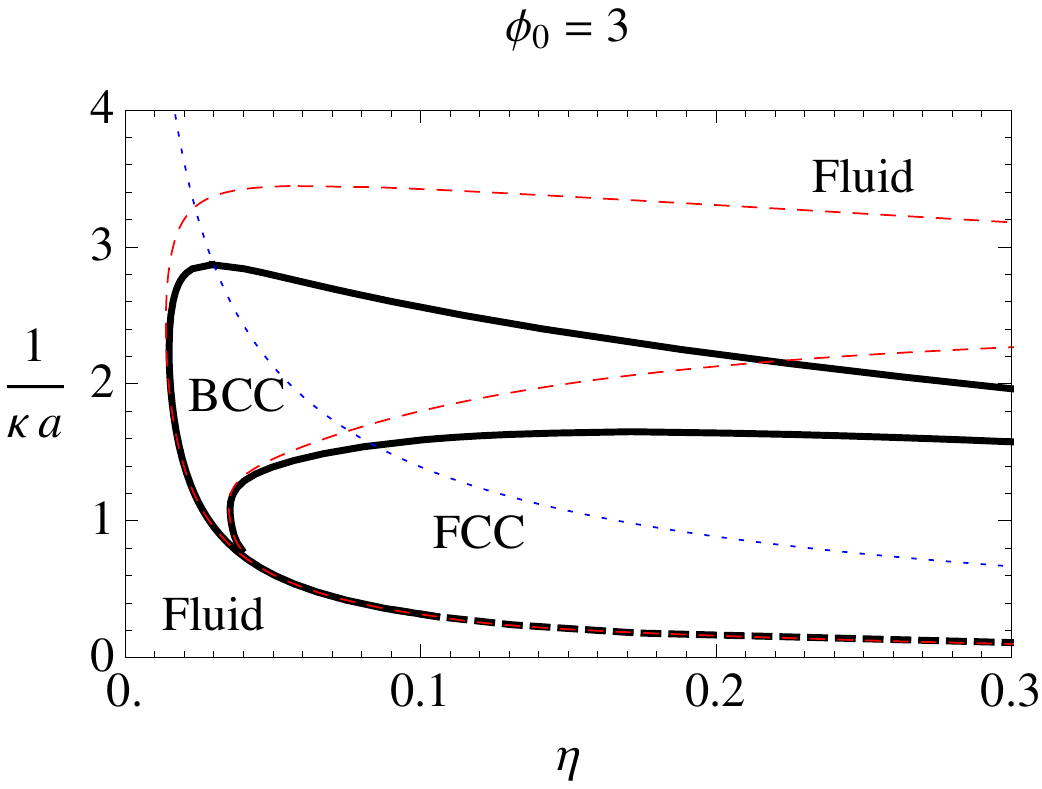}
\includegraphics[width=\figwidth]{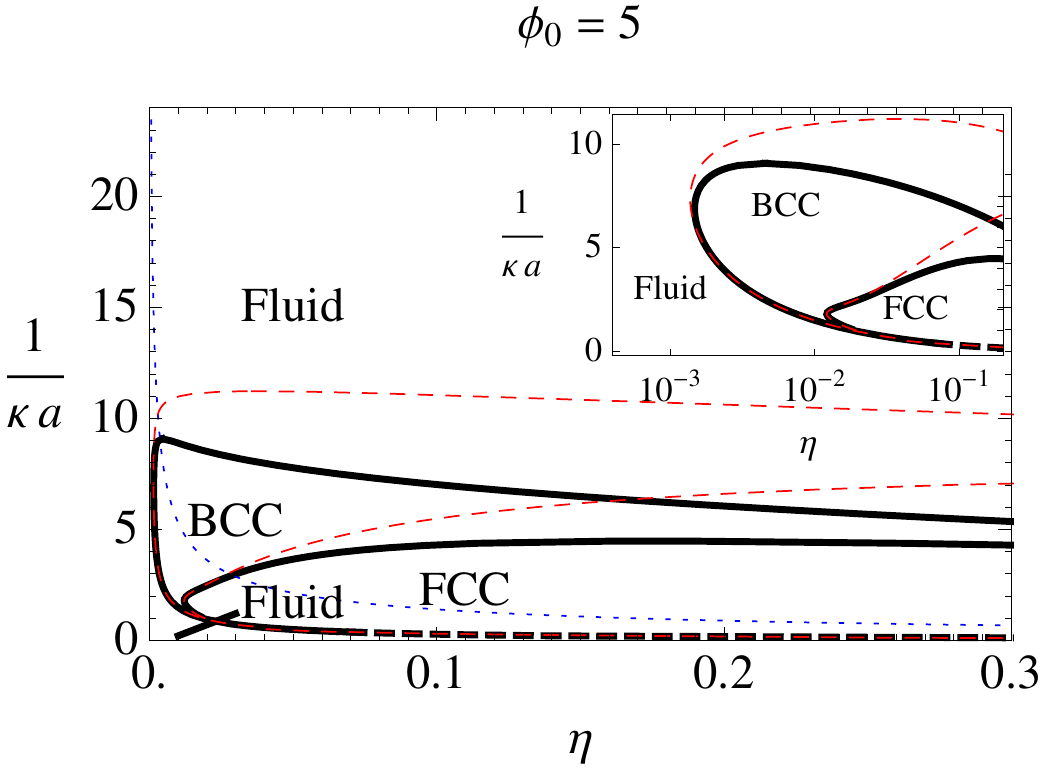}
\caption{Phase diagrams in the packing fraction-screening length ($\eta,\kappa^{-1}$) representation for constant-potential colloids (radius
$a/\lambda_B=100$) interacting with the hard-core Yukawa potential of Eq.(\ref{eq:dlvo}), for surface potentials $\phi_0 = 1, 2, 3$, and 5. The
black lines represent phase boundaries for the constant-potential model, and the red dashed lines for the association-dissociation model with the
surface potential equal to $\phi_0$ in the dilute limit. The dashed black lines indicate extrapolation of Eq.(\ref{hamaeq}) beyond its strict
regime of accuracy. The inset in the phase diagram for $\phi_0=5$ represents $\eta$ on a logarithmic scale for clarity. The labels "Fluid", "BCC",
and "FCC" denote the stable fluid, bcc, and fcc regions. We note that the very narrow fluid-fcc, fluid-bcc, and fcc-bcc coexistence regions are
just represented by single curves. The dotted blue curves represent the estimated crossover-packing fraction $\eta^*$ of Eq.(\ref{Zeta}), beyond
which $Z(\eta)<Z(0)/2$.} \label{phasediag}
\end{figure*}
\end{center}
%%%%%%%% FIG   %%%%%%%%%%%%%%%%%%%%%%%%%%%%%%%%%%%%%%%%%%%%%%%%%%%%%%%%%

The point-Yukawa system has been studied by simulation in great detail over the years,\cite{hynninen,rkg,meijer,hamaguchi} and by now it is well
known this model features a disordered fluid phase and two crystalline phases (face-centered cubic (fcc) and body-centered cubic (bcc)). Their
first-order phase boundaries are well-documented, and can accurately be described by curves in the two-dimensional $(\lambda,U)$ plane. Here, we
employ the fits for the phase boundaries of point-Yukawa particles that were presented in Ref.\cite{hynninen}, which were based on the results of
Hamaguchi {\it et al.}\cite{hamaguchi} The melting-freezing line between the bcc crystal and the fluid is accurately fitted by
\begin{eqnarray}
 \ln U &=& 4.670 - 0.04171 \lambda +0.1329 \lambda^2 -0.01043\lambda^3  \nonumber \\
           & & +\, 4.343 \cdot 10^{-4} \lambda^4 -6.924 \cdot 10^{-6} \lambda^5, \label{hamaeq}\\
           & & \mathrm{for}\,\, 0 \leq \lambda \leq 12, \nonumber
\end{eqnarray}
and the bcc-fcc transition by
\begin{eqnarray}
 \ln U &=& 97.65106 - 150.469699\lambda + 106.626405\lambda^2  \nonumber \\
           & & - 41.67136\lambda^3 + 9.639931\lambda^4 - 1.3150249\lambda^5 \nonumber\\
           & & + 0.09784811\lambda^6 - 0.00306396\lambda^7,\label{hamaeq2}\\
           & & \mathrm{for}\,\, 1.85 \leq \lambda \leq 6.8. \nonumber
\end{eqnarray}
Here we exploit these empirical relations as follows. For given dimensionless colloid radius $a/\lambda_B$ and screening constant $\kappa a$,  we
calculate $Z^*$ and $\bar{\kappa}a$ for various $\eta$ for the CSP and the AD model in the Wigner-Seitz cell, as described in the previous
section. These quantities can be used to compute the dimensionless Yukawa parameters $U$ and $\lambda$ from Eqs.(\ref{U}) and (\ref{lambda}), such
that their phase and phase-boundaries are known from Eqs.(\ref{hamaeq}) and (\ref{hamaeq2}).

\begin{center}
\begin{figure*}
\includegraphics[width=\figwidth]{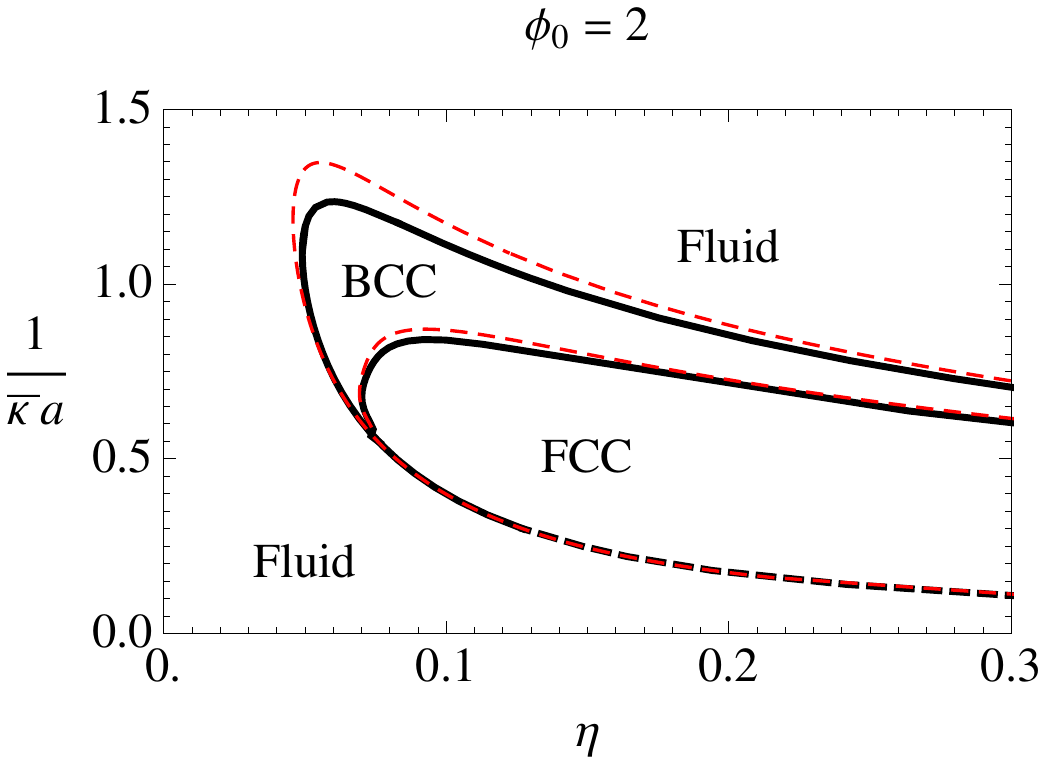}
\includegraphics[width=\figwidth]{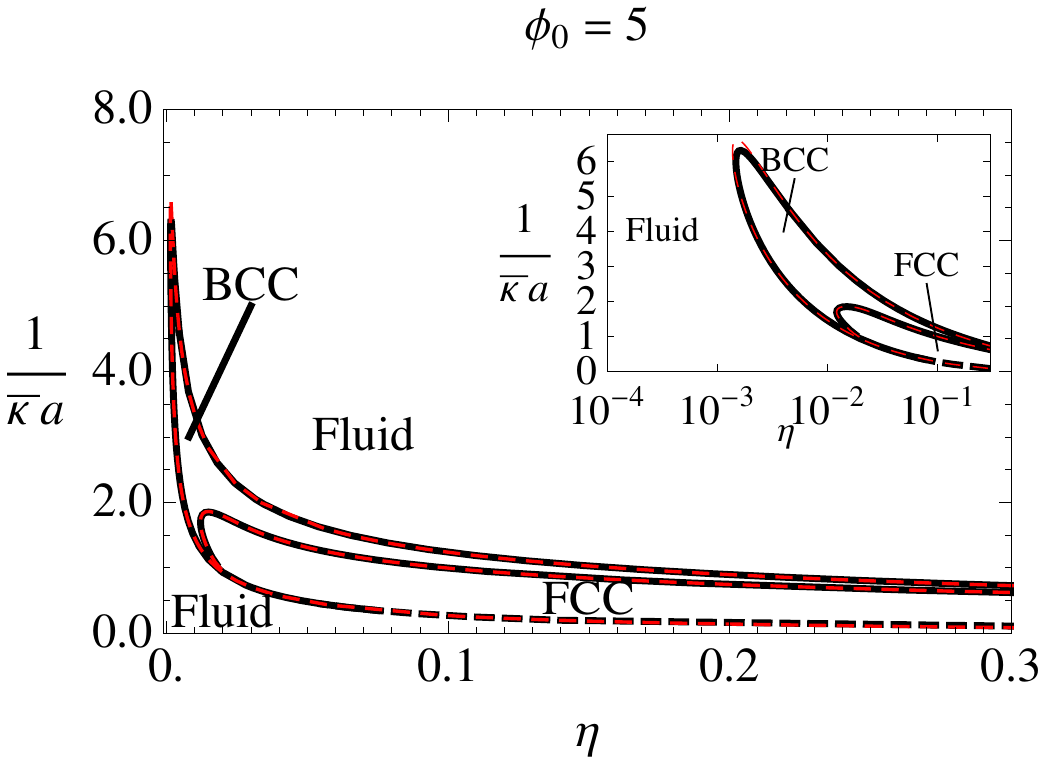}
 \caption{Phase diagrams in the packing fraction-{\em effective} screening length representation ($\eta,(\bar{\kappa}a)^{-1}$), for  $a/\lambda_B=100$,
for constant-potential colloids with (a) $\phi_0=2$ and (b) $\phi_0=5$, as well as for charge-regulated colloids. Lines, symbols, and colors as in
Fig.\ref{phasediag}.} \label{phasediagbar}
\end{figure*}
\end{center}

For $a/\lambda_B=100$, Fig.\ref{phasediag} shows the phase diagrams  that result from this point-Yukawa mapping procedure in the $(\eta,(\kappa
a)^{-1})$ representation, for the CSP model (black curves) with surface potentials (a) $\phi_0=1$, (b) $\phi_0=2$, (c) $\phi_0=3$, (d) $\phi_0=5$,
and for the corresponding AD model (red curves). The dashed lines represent the phase boundary fits of Eqs.(\ref{hamaeq}) and (\ref{hamaeq2})
outside their strict $\lambda$-regime of applicability. We restrict attention to $\eta<0.3$, as the point-Yukawa limit breaks down due to strong
excluded-volume effects at higher packing fractions.  An expected feature is the shift of the freezing curves to lower $\eta$ for higher $\phi_0$,
due to the higher (renormalized) charge and the stronger repulsions at higher $\phi_0$. Due to the higher charges in the AD model, its
crystallisation regimes (red curves) extend to somewhat lower $\eta$'s and longer screening lengths than those of the CSP model (black curves).
However, the most striking feature of all these phase diagrams is the huge extension of the fluid regime: at high {\em and} at low screening
length there is {\em no} crystalline phase at all (for $\eta<0.3$), while at some intermediate salt concentrations the crystal phases are
sandwiched in between an ordinary low-density fluid and a re-entrant fluid phase. This re-entrant fluid regime becomes more prominent with
increasing zeta-potential $\phi_0$. The underlying physics of this finite-salt and finite-$\eta$ regime where bcc and fcc crystals exist is the
{\em discharging} of the colloids with increasing $\eta$ and decreasing salt concentration: (i) although at high salt (small screening length) the
colloidal charge is high, the screened-Coulomb interaction is then so short-ranged that the system resembles a hard-sphere system that will only
crystallize at $\eta\simeq 0.5$; (ii) at low salt (long screening length) the colloidal charge is too low to have sizeable repulsions that drive
crystallisation. Only at intermediate salt and intermediate colloidal packing the charge is high enough and the screening sufficiently long-ranged
to drive crystallisation. The dotted blue curves in Fig.\ref{phasediag} represent the crossover packing fraction $\eta^*$ of Eq.(\ref{Zeta})
beyond which the colloidal charge has been reduced to less than 50 percent of its dilute-limit value.  Clearly, our expression for $\eta^*$ indeed
roughly coincides with the onset of the re-entrant fluid regime. Eq.(\ref{Zeta}) thus provides a quick guide to estimate where or whether
re-entrant melting is to be expected at all. Interestingly, there are parameter values for $\kappa a$ and $\phi_0$ (albeit in a narrow range) where a phase sequence
fluid-bcc-fcc-bcc-fluid is predicted here upon increasing the colloidal packing fraction, i.e. not only a re-entrant fluid phase but also a
re-entrant bcc phase. Moreover, for $\eta>0.5$ one expects hard-sphere freezing into an fcc (or hcp) stacking on the basis of hard-sphere
interactions, so the fcc phase is then also re-entrant.

%%%%%%%% FIG   %%%%%%%%%%%%%%%%%%%%%%%%%%%%%%%%%%%%%%%%%%%%%%%%%%%%%%%%%
\begin{center}
\begin{figure*}
\includegraphics[width=\figwidth]{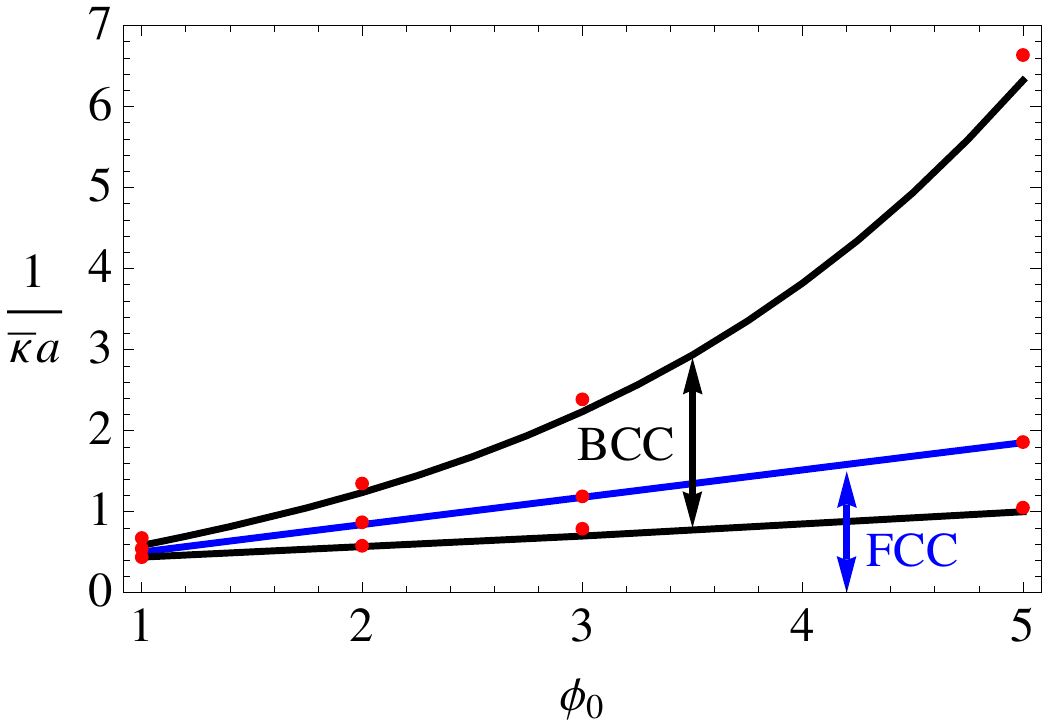}
\caption{Maximum and minimum effective screening lengths where bcc and fcc can be found, as a function of the surface potential, assuming a
constant surface potential for $a/\lambda_B=100$. The bcc regime is in between the two black lines, and the fcc regime below the blue line. The
red points indicate the results from the AD model.} \label{bccline}
\end{figure*}
\end{center}
%%%%%%%% FIG   %%%%%%%%%%%%%%%%%%%%%%%%%%%%%%%%%%%%%%%%%%%%%%%%%%%%%%%%%

Experimentally it is not always possible or convenient to characterise the screening in terms of the  Debye length $\kappa^{-1}$ of the
(hypothetical) reservoir with which the suspension would be in osmotic equilibrium. Rather, one often measures the effective (actual) screening
length $\bar{\kappa}$ in the suspension of interest. For this reason we replot in Fig.\ref{phasediagbar} the phase diagrams for $\phi_0=2$ and
$\phi_0=5$ of Fig.\ref{phasediag}, but now in the $(\eta,(\bar{\kappa}a)^{-1})$ representation. Interestingly, the CSP and AD model are now much
closer together, and the re-entrant fluid phase appears even more pronounced in this representation.

In order to quantify in which finite salt-concentration regime bcc and fcc crystals are expected in a colloidal concentration series $0<\eta<0.3$,
we analyse the maximum and minimum values of $\bar{\kappa}a$ at which these two crystal phases can exist, as a function of the zeta-potential
$\phi_0$, for $a/\lambda_B=100$. Fig. \ref{bccline} shows the resulting screening-length regimes, both for bcc (black curves) and fcc (blue
curve), where the lowest screening length for fcc crystals is set to zero because of the hard-sphere freezing into fcc at $\eta=0.5$ even for
$1/\kappa a\rightarrow 0$
---of course we only restricted attention to $\eta<0.3$ until now so strictly speaking also the fcc phase should have had a nonvanishing lower
bound. Nevertheless, despite this small inconsistency, Fig.\ref{bccline} clearly shows not only that a larger zeta-potential gives rise to a
larger crystal regime, but also that for all $\phi_0$ there is a limiting screening length beyond which neither fcc nor bcc crystals can exist,
both for the CSP and the AD model.

\begin{center}
\begin{figure*}
\includegraphics[width=\figwidth]{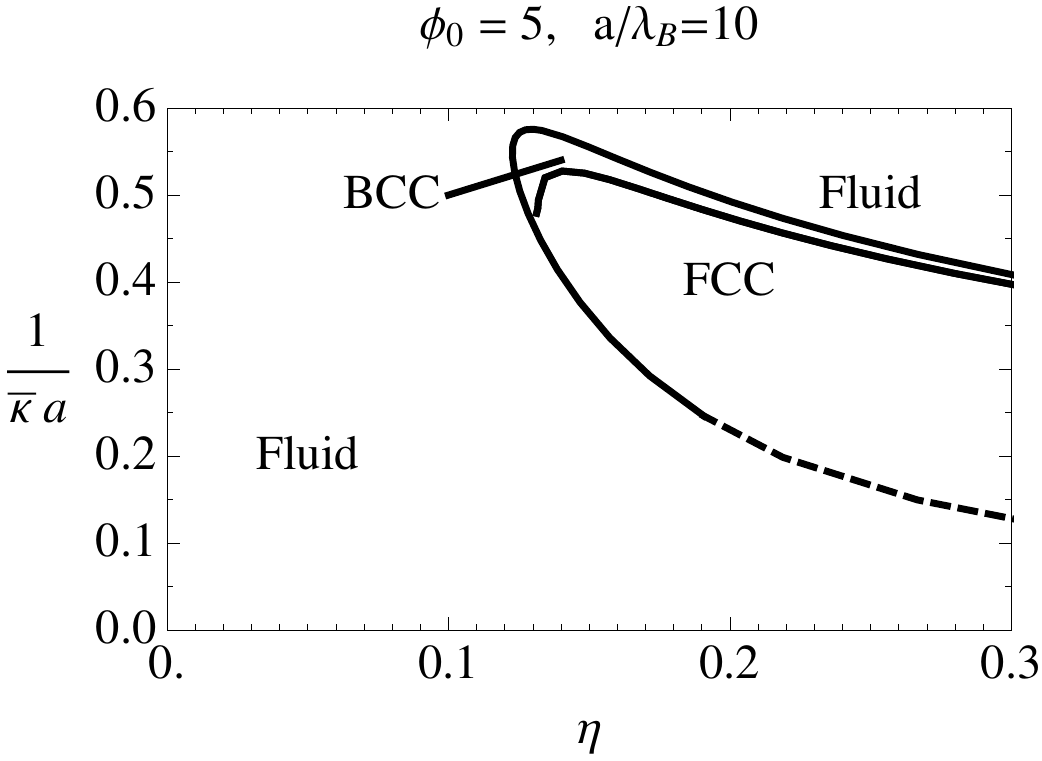}
\includegraphics[width=\figwidth]{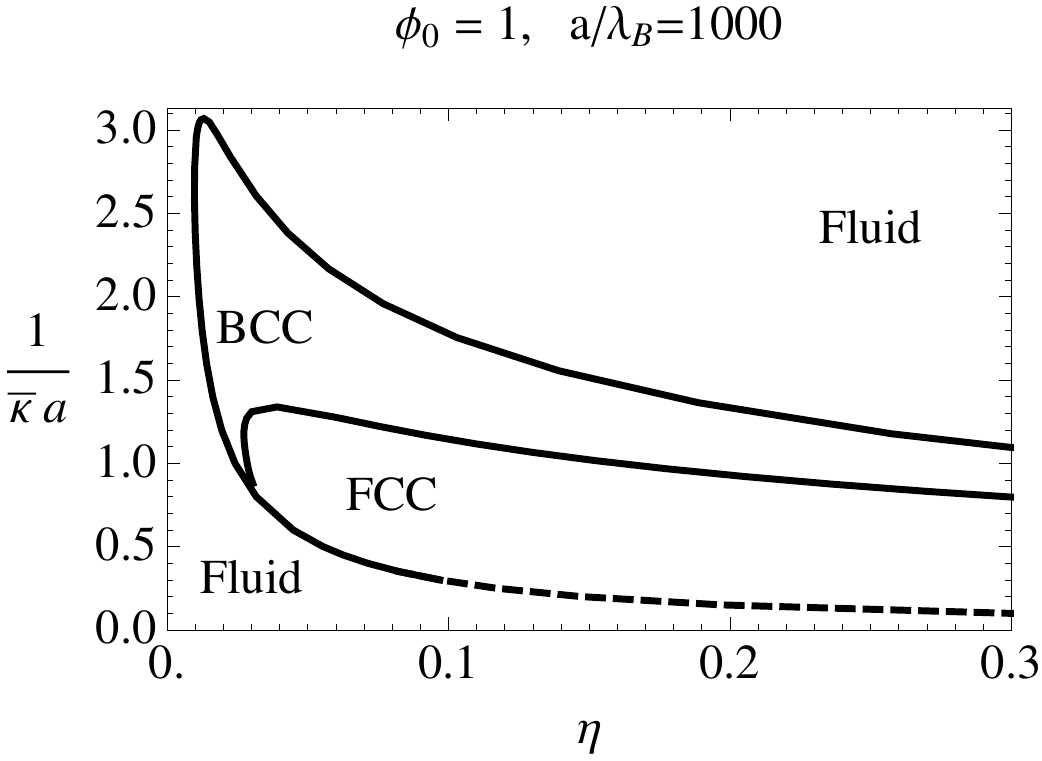}
 \caption{Phase diagrams in the packing fraction-{\em effective} screening length representation ($\eta,(\bar{\kappa}a)^{-1}$),
for constant-potential colloids with (a) $\phi_0=5$ for  $a/\lambda_B=10$ and (b) $\phi_0=1$ and for  $a/\lambda_B=1000$. Lines and symbols as in
Fig.\ref{phasediag}.} \label{phasediagbarlz}
\end{figure*}
\end{center}

So far we focussed on $a/\lambda_B=100$, which in aqueous suspensions corresponds to a colloidal radius of about 70nm. However the colloidal size
regime can easily be a factor 10 larger or smaller, and for that reason we also consider the CSP model for $a/\lambda_B=10$ and 1000. In
Fig.\ref{phasediagbarlz} we show the phase diagrams for the smaller colloids with $\phi_0=5$ in (a), and for larger colloids with $\phi_0=1$ in
(b). Interestingly, they resemble those for $a/\lambda_B=100$ shown in Fig.\ref{phasediagbar}(a) and (b) for $\phi_0=2$ and 5, respectively, but
the smaller colloids require a much higher potential while the larger ones only need a lower potential to obtain phase diagrams similar to those
for $a/\lambda_B=100$ ---the phase diagram for $a/\lambda_B=10$ at $\phi_0=1$ does not show a crystal phase at all for $\eta<0.3$. In other words,
given that $\phi_0=5$ is a rather high potential that may be difficult to achieve in reality while $\phi_0=1$ is frequently occurring, one
concludes that re-entrant melting occurs in the largest salt-concentration regime (and is hence easiest observable by tuning the salt) for larger
colloids.

\section{Summary and conclusions}
Within a Wigner-Seitz cell model we have calculated the bare charge $Z$, the renormalized charge $Z^*$, and the effective screening length
$\bar{\kappa}^{-1}$ of colloidal spheres at a constant zeta-potential $\phi_0$. We find from numerical solutions of the nonlinear
Poisson-Boltzmann equation that these constant-potential colloids discharge with increasing packing fraction and ionic screening length, in fair
agreement with analytical estimates for the dilute-limit charge $Z_0$ in Eq.(\ref{Z0}) and the typical crossover packing fraction $\eta^*$ given
in Eq.(\ref{Zeta}). We also show that the constant-potential assumption is a reasonably accurate description of charge regulation by an ionic
association-dissociation equilibrium on the colloidal surface. We use our nonlinear calculations of $Z^*$ and $\bar{\kappa}$ to determine the
effective screened-Coulomb interactions between the colloids at a given state point, and we calculate the phase diagram for various
zeta-potentials by a mapping onto empirical fits of simulated phase diagrams of point-Yukawa fluids. This reveals a very limited regime of bcc and
fcc crystals: in order to form crystals, the charge is only high enough and the repulsions only long-ranged enough in a finite intermediate regime
of packing fraction and salt concentrations; at high $\eta$ or low salt the spheres discharge too much, and at high salt the repulsions are too
short-ranged to stabilise crystals. In the salt-regime where crystals can exist, the discharging mechanism gives rise to re-entrant phase
behaviour, with phase sequences fluid-bcc-fluid and even fluid-bcc-fcc-bcc-fluid upon increasing the colloid concentration from extremely dilute
to $\eta=0.3$.

The phase behaviour of constant-potential or charge-regulated colloids as reported here is quite different from that of constant-charge colloids,
for which the pairwise repulsions do not weaken with increasing volume fraction or decreasing salt concentration. As a consequence constant-charge
colloids have a much larger parameter-regime where crystals exist, and do not show the re-entrant behaviour.\cite{rkg,meijer,hamaguchi,hynninen}
The most direct comparison is to be made with the constant-charge phase diagrams of Fig.2 and Fig.4 of Ref.\cite{hynninen}, where the charge is
fixed such that the surface potential at infinite dilution corresponds to $\phi_0\simeq 1$ and 2, respectively. Our theoretical findings can thus
be used to gain insight into the colloidal charging mechanism by studying colloidal crystallisation regimes as a function of packing fraction and
salt concentration.

\section{Acknowledgement}
Financial support of an NWO-ECHO and an NWO-VICI grant is acknowledged.

\section*{Appendix}
Although it is numerically straightforward to solve the nonlinear PB equation (\ref{pb1}) with BC's (\ref{bc1})
and (\ref{bc2}) in a spherical Wigner-Seitz cell of radius $R$, it may also be  convenient to have analytic
results that allow for quick estimates of the (order of) magnitude of the colloidal charge $Z$. A standard
approach is to linearise the $\sinh\phi(r)$ term of Eq.(\ref{pb1}), e.g. with $\phi(r)-\phi(R)$ as the small
expansion parameter. The resulting solution is then of the form $\phi(r)=A\exp(-\bar{\kappa}r)/r +B
\exp(\bar{\kappa}r)/r +C$, with $\bar{\kappa}$ defined in Eq.(\ref{kappabar}), $C=\phi(R)-\tanh\phi(R)$, and with
integration constants $A$ and $B$ fixed by the two BC's. The algebra involved is, however, not very transparent.

A considerable simplification is achieved if we consider the so-called Jellium model, in which the central
colloidal sphere is no longer considered to be surrounded by only cations and anions in a finite cell, but instead
by cations, anions {\em and} other colloids with charge $Z$ (to be determined).\cite{trizac1, trizac2, trizac3} A nonlinear PB
equation and BC's can then be written, for $r\geq a$,
\begin{eqnarray}
\phi''(r)+\frac{2}{r}\phi'(r)&=&\kappa^2\sinh\phi(r)-4\pi\lambda_B Zn;\label{pb2}\\ % , \hspace{1mm}r\geq a
\phi(a)&=&\phi_0; \label{bc3}\\
\phi'(\infty)&=&0,\label{bc4}
\end{eqnarray}
where it is assumed that the 'other' colloids are distributed homogeneously with density $n$. From this one
derives directly that the asymptotic potential is given by
\begin{eqnarray} \sinh\phi(\infty)=\frac{4\pi \lambda_B Z n}{\kappa^2}=\frac{3\eta
(Z\lambda_B/a)}{(\kappa a)^2}.\label{phid}
\end{eqnarray}
Now linearising $\sinh\phi(r)$ with $\phi(r)-\phi(\infty)$ as the small expansion parameter gives rise to the
solution
\begin{equation}\label{linphi}
\phi(r)=\phi(\infty)+ (\phi_0-\phi(\infty))\frac{\exp\big(-\tilde{\kappa}(r-a)\big)}{r/a},
\end{equation}
where the effective screening length $\tilde{\kappa}^{-1}$ is defined by
\begin{equation}
\tilde{\kappa}=\kappa \sqrt{\cosh\phi(\infty)}. \label{kappatilde}
\end{equation}
We note that the average ion concentrations in the system, within the present linearisation scheme, is given by
$c_{\pm}=\rho_s\exp(\mp\phi(\infty))$, such that the corresponding screening length $\tilde{\kappa}^{-1}$ is given
by $\tilde{\kappa}^2=4\pi\lambda_B(c_++c_-)$. In other words, the effective screening length $\tilde{\kappa}$ and
the asymptotic potential $\phi(\infty)$ of this jellium model play exactly the same role as $\bar{\kappa}$ and
$\phi(R)$ that we introduced before in the Wigner-Seitz cell. In particular, $\bar{\kappa}^{-1}$ and
$\tilde{\kappa}^{-1}$ can be seen as the {\em actual} screening length in the suspension (in contrast to the
screening length $\kappa^{-1}$ of the ion reservoir).

From Eq.(\ref{linphi}) the colloidal charge $Z$ follows, using Eq.(\ref{Zws}), as the solution of the
transcendental equation
\begin{equation}
\frac{Z\lambda_B}{a}=(\phi_0-\phi(\infty))(1+\tilde{\kappa}a),\label{sc}
\end{equation}
where one should realise that both $\phi(\infty)$ and $\tilde{\kappa}$ depend on $Z\lambda_B/a$ through
Eqs.(\ref{phid}) and (\ref{kappatilde}). It is possible to solve Eq.(\ref{sc}) explicitly in the dilute limit. For
$\eta=0$ one finds $\phi(\infty)=0$ from Eq.(\ref{phid}), and hence $Z=Z_0$ given by Eq.(\ref{Z0}). For finite but
low-enough $\eta$ for which $\phi(\infty)\ll 1$ one can ignore ${\cal O}(\eta^2)$ contributions, such that
$\sinh\phi(\infty)\simeq \phi(\infty)$ and $\cosh\phi(\infty)\simeq 1$, to find Eq.(\ref{Zeta}) from the
self-consistency condition Eq.(\ref{sc}).

\end{document}